# Assessing Cybersecurity Risks and Traffic Impact in Connected Autonomous Vehicles

Saurav Silwal[1*], Lu Gao, Ph.D.[2], Yunpeng Zhang, Ph.D.[3], Ahmed Senouci, Ph.D.[4], Yi-Lung Mo, Ph.D., P.E.[5]

[1]Department of Construction Management and Department of Civil and Environmental Engineering, University of Houston, Houston, TX 77204; e-mail: ssilwal@cougarnet.uh.edu
[2]Department of Construction Management, University of Houston, Houston, TX 77204-4020; email: lgao5@central.uh.edu
[3]Department of Information Science Technology, University of Houston, Sugar Land, TX 77479; email: yzhang119@uh.edu
[4]Department of Construction Management, University of Houston, Houston, TX 77204-4020; email: asenouci@uh.edu
[5]Department of Civil and Environmental Engineering, University of Houston, Houston, TX 77204-4003; email: yilungmo@central.uh.edu

**ABSTRACT**

Given the promising future of autonomous vehicles, it is foreseeable that self-driving cars will soon emerge as the predominant mode of transportation. While autonomous vehicles offer enhanced efficiency, they remain vulnerable to external attacks. In this research, we sought to investigate the potential impact of cyberattacks on traffic patterns. To achieve this, we conducted simulations where cyberattacks were simulated on connected vehicles by disseminating false information to either a single vehicle or vehicle platoons. The primary objective of this research is to assess the cybersecurity challenges confronting connected and automated vehicles and propose practical solutions to minimize the adverse effects of malicious external information. In the simulation, we have implemented an innovative car-following model for the simulation of connected self-driving vehicles. This model continually monitors data received from preceding vehicles and optimizes various actions, such as acceleration, and deceleration, with the aim of maximizing overall traffic efficiency and safety.

## INTRODUCTION

Technological advancements consistently bring forth new innovations across diverse fields. In recent years, researchers have made significant strides by introducing semi-autonomous vehicles, fulfilling humanity's longstanding aspiration for autonomous travel. Moreover, the escalating issues of traffic congestion and accidents highlight a pressing need for innovative remedies, propelling this sector toward further evolution. A compelling aspect of autonomous vehicles is Connected Autonomous Vehicles (CAVs), where vehicles interact on the road, exchanging critical data to enable autonomous driving. Bajpai (2016) explores the promising effects of CAVs, demonstrating their potential to mitigate congestion, enhance safety, and reduce fuel usage.

CAVs employ adaptive cruise control (ACC) and its advanced iteration, connected adaptive cruise control (CACC). ACC is a driving assistant and functions using the vehicle's sensors to automatically adjust its speed relative to the preceding vehicle, whereas CACC leverages vehicle-to-vehicle(V2V) communication, sharing data like the vehicle's speed, location, and activity, facilitating a more synchronized and seamless driving experience (Van Arem et.al 2006). Over the years, researchers have examined how implementing ACC or CACC in vehicles enhances traffic dynamics. VanderWerf et.al. (2002) employed Monte Carlo simulations to conduct a quantitative assessment of the effectiveness of ACC and concluded that ACC has the potential to enhance highway traffic throughput, particularly when CACC is integrated into vehicles. Similarly, Ploeg et.al. (2011) conducted experiments involving six CACC-equipped vehicles, affirming that CACC improves traffic flow.

CAVs present numerous promising benefits for the future of transportation, yet they also bring forth significant vulnerabilities. Among these vulnerabilities, one of the primary concerns is the susceptibility to cyber-attacks (Gao et al., 2025). The inclusion of wireless communication between vehicles exposes them to various types of attacks that could potentially cause harm or damage. Given the pervasive use of wireless technology in our daily lives, we're aware of the damaging potential of cyber threats. It's crucial to recognize that vehicles equipped with wireless communication are susceptible to attacks from individuals with malicious intentions. Addressing such threats requires a proactive approach: first identifying potential attack vectors and understanding their possible impacts. Armed with this knowledge, efforts can be directed toward developing defenses to safeguard these vehicles against potential cyber-attacks. Yağdereli et.al. (2015) explored the evolving landscape of autonomous transportation modes, detailing diverse categories of cyber threats: passive and active attacks. They offered potential mitigation strategies while emphasizing the pressing need for an update to the Control Area Network (CAN) standard, which currently stands as outdated in the light of emerging cyber threats. Similarly, Parkinson et.al. (2017) conducted a comprehensive analysis of existing literature in this domain, revealing a significant gap in knowledge within the field and the need for extensive research endeavors to fill this void and fortify cybersecurity measures.

Numerous researchers have proposed diverse approaches to comprehensively grasp the cyber threats faced. They've delineated different types of cyber-attacks, ranging from sending false messages, and replaying stored messages, to manipulating acceleration or deceleration, among others (Lebaku et al., 2025). By employing different car-following models, researchers aimed to simulate traffic flow scenarios and scrutinize vehicle behavior when subjected to these cyber-attacks, whether it's an individual vehicle or multiple vehicles affected. Wang et.al. (2018), for instance, utilized an extended optimal car-following model. Their study focused on analyzing the impacts of cyber-attacks that altered spacing and velocity, evaluating how these disruptions impacted traffic flow stability. Similarly, Sun et.al. (2023) conducted microsimulation modeling to examine time-delay and disturbance attacks at freeways and unsignalized intersections, uncovering the risk posed by these attacks, leading to collisions, congestion, and reduced road capacity. Furthermore, Khattak et.al. (2021) modeled multiple CAV platoons and simulated lane-changing dynamics. They found that cyber-attacks significantly jeopardized platoon stability and safety, with lane-changing conflicts posing greater challenges than rear-end collisions. In an extensive study examining the impact of cyber-attacks on CAV platoon stability, Wang et.al. (2023) incorporated sensors along with an augmented state extended Kalman filter to detect and filter anomalous readings. They conducted sensitivity analysis and obtained a critical detection sensitivity, which ensures the effect of detected faulty reading are not amplified as information is relayed to subsequent vehicles, thereby maintaining platoon stability.

The journey toward achieving complete autonomy in vehicles is still in its early stages, and conducting extensive research into the potential weakness of autonomous vehicles remains crucial. Current research on the cybersecurity threats to CAVs is somewhat limited. This paper seeks to add to this body of knowledge. Our study involves the simulation of a car-following model where we've introduced attacks such as injecting false messages, replaying stored data, and manipulating acceleration to assess their impact on the traffic flow.

## METHODOLOGY

This section presents the car following model utilized for simulating traffic flow. Following the establishment of this traffic model, cyber-attacks were instigated to observe their impact on the traffic flow.

### Car Following Model

The car-following model is the mathematical representation of the behavior of vehicles in a traffic stream. Pipes (1953) was the pioneer in introducing a car-following model to describe the characteristics of vehicles moving in a stream. The model described how a vehicle reacts to the changes in the vehicle in front of it. The development of a car-following model as a mathematical representation has made it possible to conduct stability analysis and made it possible to gauge the safety and traffic disruptions and propel the study toward autonomy (RW Rothery, 1992). At its core, the basic car-following model operates on the principle that the vehicle's acceleration, i.e., the response is determined by the stimulus, which is the velocity difference between the vehicle and its leader (Chandler et.al. 1958). Over time researchers have progressively refined the car-following model, striving for greater realism to mirror real-life scenarios in traffic. Treiber et.al (2000) proposed a new model called the intelligent driver model (IDM) that is an improvement on the classical models. This new model demonstrates the capacity to portray realistic vehicle behavior within single-lane traffic scenarios by incorporating relevant parameters.

### Intelligent Driver Model

The IDM computes the subject vehicle's acceleration utilizing data from the lead vehicle, as expressed in the following equation:

$$\dot{v}_n(t) = a\left(1 - \left(\frac{v_n(t)}{v_0}\right)^\delta - \left(\frac{s*(v_n(t),\Delta v_n(t))}{s_n(t)}\right)^2\right) \quad (1)$$

$$s*(v_n(t), \Delta v_n(t)) = s_0 + Tv_n(t) + \frac{v_n(t)*\Delta v_n(t)}{2\sqrt{ab}} \quad (2)$$

where, $s_n(t) = x_{n-1} - x_n - l$, where $n-1$ represents the preceding vehicle, $n$ represents the subject vehicle, $l$ is the length of the vehicle, and $\Delta v_n(t) = v_n(t) - v_{n-1}(t)$.

Here, $v_0, s_0, a, b$ and $T$ are model parameters.
$v_0$ = desired velocity, the velocity of the vehicle in free traffic
$s_0$ = minimum spacing between the vehicles

$a$ = maximum acceleration
$b$ = maximum comfortable deceleration
$T$ = minimum possible time to vehicle in front
$\delta$ = acceleration exponent

The IDM has been widely employed in various studies to simulate longitudinal vehicle dynamics. Despite its effectiveness as a model, the classical IDM's limitation lies in its focus solely on the influence of the immediately preceding vehicle. Previous studies have generalized the IDM to accommodate connected vehicle settings. For example, Wang et.al (2020) introduced a new approach within the classical IDM by integrating weighing coefficients for spacing and velocity differences, thus accommodating the impact of multiple leading vehicles. In this method, the acceleration of the target vehicle is influenced by the gap and speed difference relative to the vehicles in front of it. Although using weights to determine the degree influence of preceding vehicles is good, this approach fails to represent a true-to-life situation. According to this model, if the nearest leading vehicle is abruptly halted, the subject vehicle will decelerate but, due to information from other vehicles ahead, it will not come to a complete stop, eventually leading to a collision. Thus, in this study, we developed a novel approach to extend the IDM to allow the subject vehicle to gather velocity and position data from preceding vehicles. Leveraging this information, the subject vehicle utilizes IDM to calculate acceleration with respect to each of these leading vehicles and select the one with the safest consequence. Furthermore, for simplicity, in this study reaction time was not included.

$$\dot{v}_n(t) = \min(\dot{v}_n^{n-1}(t), \dot{v}_n^{n-2}(t), \dots, \dot{v}_n^{n-m}(t)) \qquad (3)$$

where,

$\dot{v}_n^j(t)$ is the acceleration calculated using Equation (1) and (2) by assuming vehicle $j$ is the vehicle right in front of vehicle $n$. Figure 1 shows a simplified representation of connected automated vehicles setting considered in this paper. On a single-lane road, multiple autonomous vehicles are in motion. These vehicles have established a wireless network to share information about their speed and position, assisting each vehicle in deciding its next move. This study focuses exclusively on the scenario where vehicles communicate only with those ahead of it.

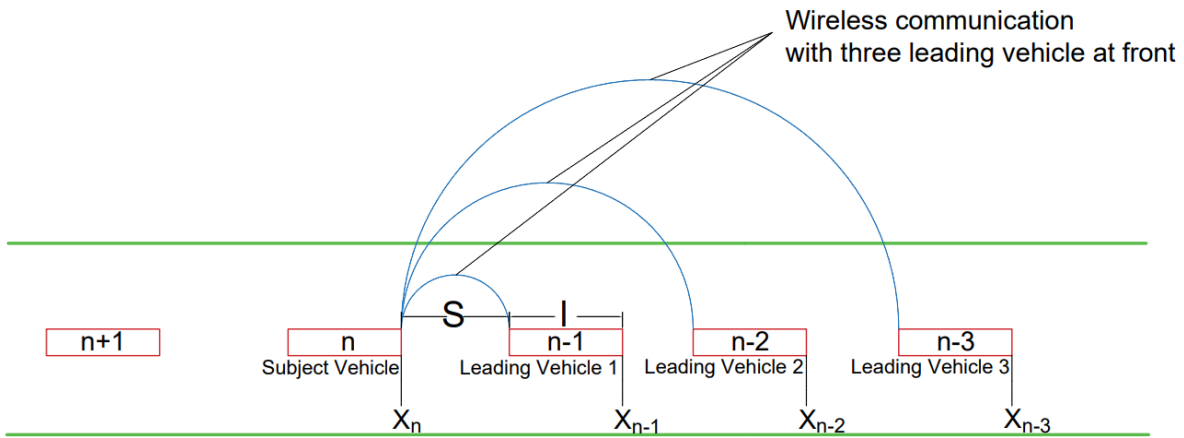

**Figure 1. Typical Representation of CAV**

**Cyberattacks**

CACC-equipped vehicles engage in wireless communication among themselves. However, in the event of a compromise or cyber-attack on this communication network, vehicles still rely on in-vehicle sensors to gather essential information. Our study aims to investigate the ramifications of cyber-attacks by exploring a scenario where both wireless communication and sensors are compromised. This worst-case scenario leaves the vehicle highly susceptible to malicious attacks, enabling us to evaluate the severe impact such vulnerabilities might cause. The cyber-attacks that have been implemented in this study are as follows.

i. False Message: The attacked vehicle is fed erroneous information regarding the leader vehicle's position and velocity.
ii. Platoon Leader Identity Attack: This cyberattack targets a vehicle, causing it to no longer recognize any vehicles positioned ahead of it, consequently perceiving itself as the leading vehicle within the platoon.
iii. Acceleration Manipulation: This attack involves replacing the acceleration value calculated by Equation (3) with an alternative value.
iv. Multiple Vehicle Attack: Simultaneously targeting multiple vehicles by providing them with misleading or false information.

These attacks cause disturbances in traffic flow, as demonstrated in the subsequent results section. Rear-end collisions or travel delays have been identified as key indicators of system failure.

**Simulation Setup**

Python programming language was used to create a simulation platform. A total of nine vehicles were defined to set up a vehicular platoon on a single-lane road. Overtaking or lane-changing behaviors were not considered for this study. Initially, specific positions and velocities were assigned to each vehicle. However, during simulation, the vehicles adapted their characteristics following Equation (3). Table 1 outlines the parameters we employed and their respective values for the study.

**Table 1 Simulation Parameters**

| Parameters | Values |
|---|---|
| Number of lanes | 1 |
| Length of road segment | 1000 m |
| Desired speed ($v_0$) | 10 m/s |
| Length of vehicles (l) | 5 m |
| Minimum gap ($s_0$) | 2 m |
| Headway time (T) | 1.5 sec |
| Maximum acceleration (a) | 0.73 m/s$^2$ |
| Maximum deceleration (b) | 1.67 m/s$^2$ |
| Acceleration exponent ($\delta$) | 4 |

To simplify our study, the vehicle's communication range was limited to a maximum of three vehicles ahead. This decision was based on the understanding that distant vehicles exert less influence on the decision made by the subject vehicle. This range can be easily extended to more vehicles if needed.

## RESULT AND DISCUSSION

Following the development of a simulation platform, various cyber-attacks were initiated to assess their influence on traffic flow. Figure 2 shows the CAV platoon in the absence of any attacks. This figure serves as a basis in comparing cyberattack scenarios. In the absence of an attack, the graphs exhibit smoothness, showcasing the vehicles' uninterrupted travel. The graph illustrating car positions over time demonstrates their smooth functioning without any deviations. Initially, the cars accelerate, then enter a phase of stability. The graphs of velocity and acceleration against simulation time illustrate this behavior. As the lead vehicle approaches the road's end, it slows down, causing the vehicles behind it to also decelerate and reduce their speed. The graph showing the gap to the leading vehicle over time represents the changing distance between each vehicle and the one in front of it over time. The blue line on this graph shows the gap perceived by the first vehicle, which initially is large due to the open road ahead but decreases as the vehicle progresses and reaches the end of the road. This decreasing blue line represent this trend.

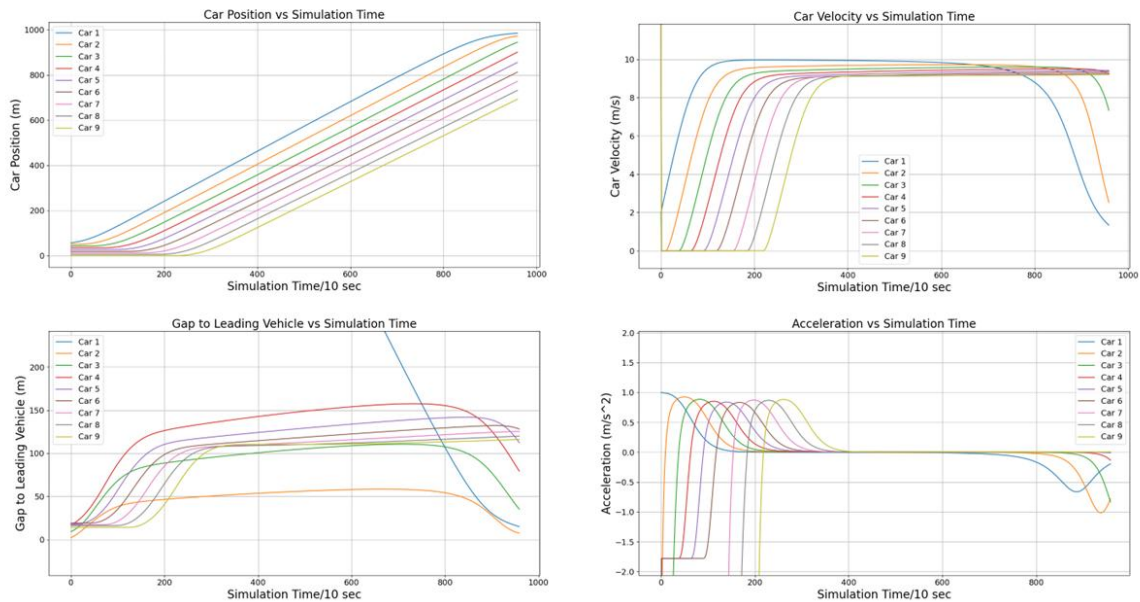

**Figure 2. CAV Platoon without Cyber Attack**

### Attack Type: False Message

In this attack scenario, the fifth car within the platoon received two types of false messages. Initially, it received erroneous data related to spacing, indicating that the three vehicles it was meant to follow were positioned 80 meters farther ahead than their actual respective positions. This deceptive information led the targeted vehicle to perceive a larger gap in front, prompting it to accelerate. The attack commenced 10 seconds after the simulation began and persisted for 75 seconds in total. Within the attack timeframe, a rear-end collision occurred between vehicle no. 5 and vehicle no. 4 at 63 seconds into the simulation. Figure 3 depicts the disruption in traffic dynamics caused by this attack.

Another attack was aimed at manipulating the speed of the leading vehicles, specifically targeting the fifth vehicle once again. This attack commenced 10 seconds after the simulation began and persisted for 70 seconds. The vehicle was compromised to perceive its lead vehicles' velocities as tenfold less than their actual speed, resulting in a significant slowdown. The sixth vehicle, trailing the fifth vehicle was examined. Under normal circumstances, this vehicle should have reached the 600-meter mark within 75.3 seconds from the start of the simulation. However, due to the attack, a delay of 3.87 seconds was incurred in reaching the same position. Figure 4 illustrates the changes in the traffic flow under this type of attack.

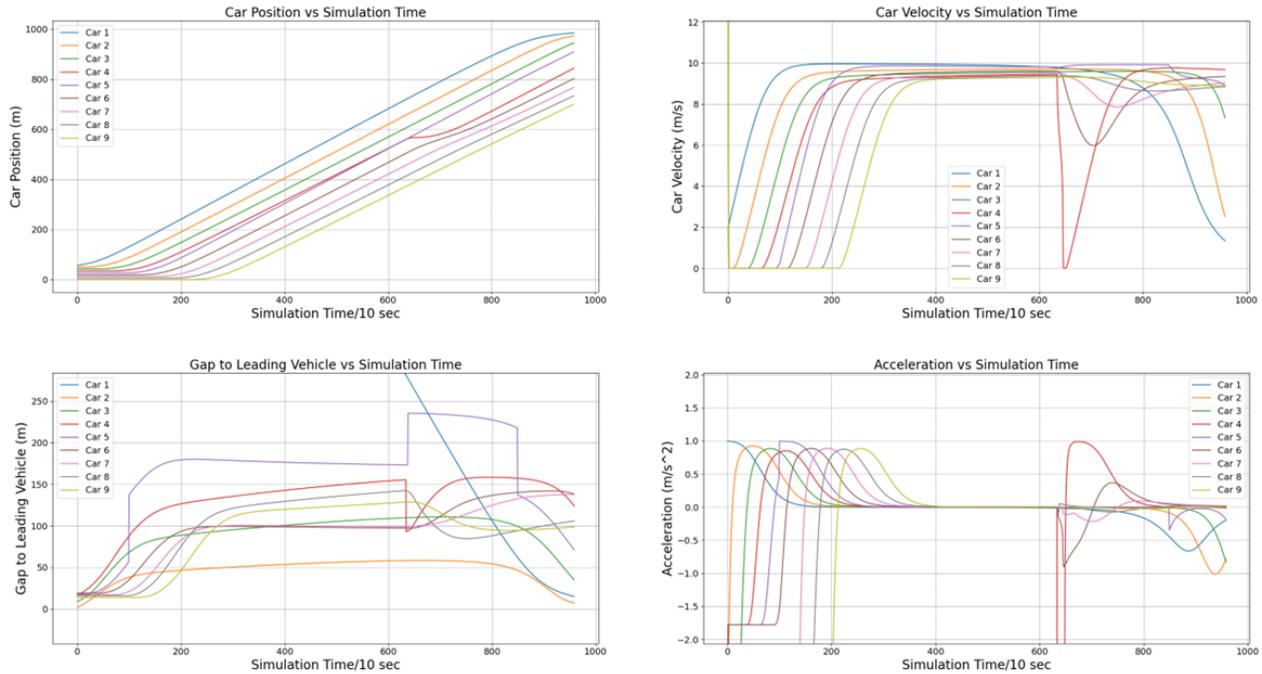

**Figure 3. CAV Platoon under Manipulated Spacing Data**

**Attack Type: Platoon Leader Identity Attack**
In this attack scenario, within the CAV platoon, the fifth vehicle was selected, and its communication was tampered with, causing a disruption in its connectivity with the preceding vehicles. The vehicle's perception was manipulated to disregard any vehicles ahead, prompting it to assume the role of the platoon leader. Consequently, the vehicle interpreted an open road ahead, initiating an acceleration that led to a rear-end collision with vehicle no. 4 at 53.4 seconds into the simulation. Figure 5 demonstrates the resultant alterations in traffic flow that ensued once the attack commenced 10 seconds into the simulation and lasted for 70 seconds.

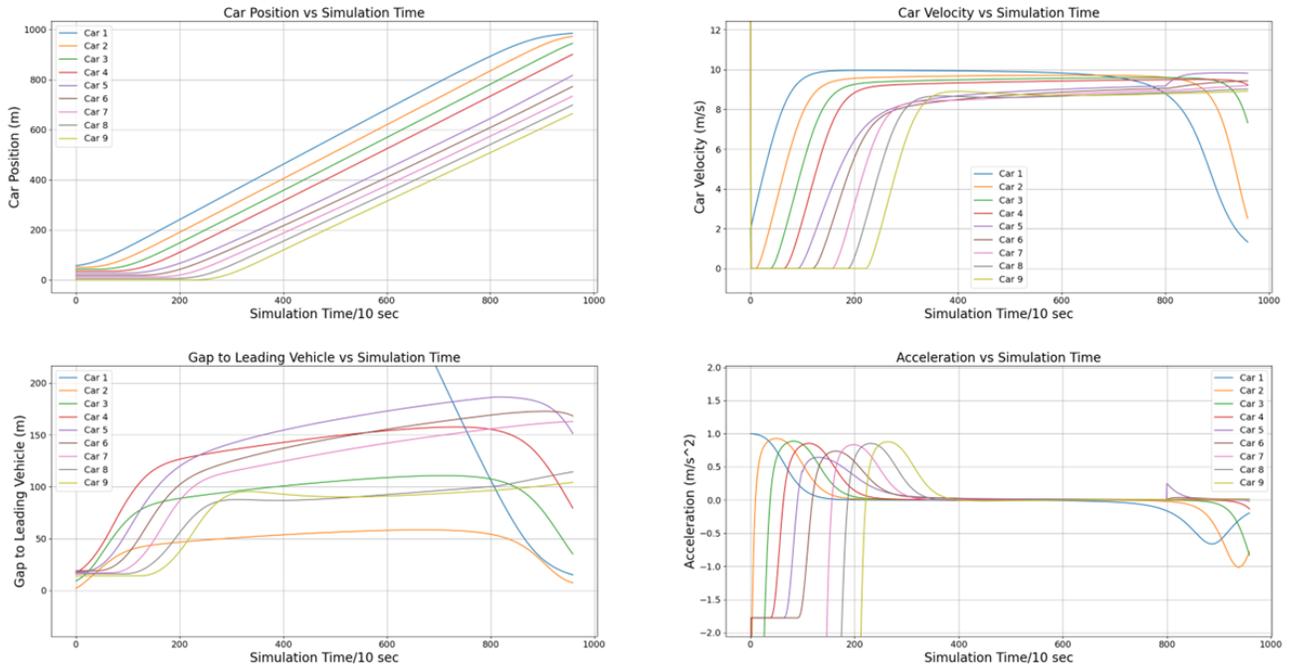

**Figure 4. CAV Platoon under Manipulated Velocity Data**

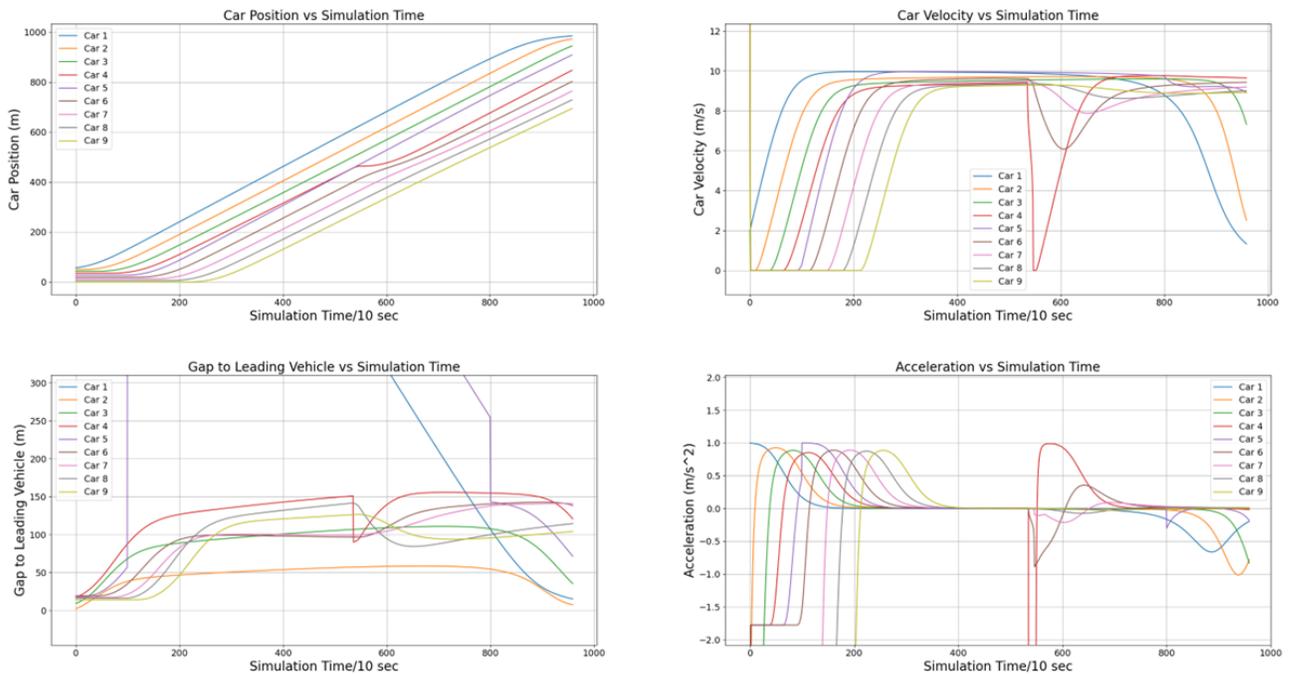

**Figure 5. CAV Platoon with Platoon Leader Identity Attack**

**Attack Type: Acceleration Manipulation**
In this scenario, at the 30-second mark within the simulation, an attack was initiated which disrupted vehicle number 5's expected acceleration, introducing a new value. This alteration in acceleration is depicted in Figure 6, displaying a sudden decline. This disruptive attack persisted for 10 seconds, enforcing a deceleration of $2m/s^2$. Consequently, the targeted vehicle experienced an abrupt slowdown, causing all the subsequent vehicles behind it to also slow down. A travel delay became apparent, as under normal conditions, vehicle number 6 would have reached the 600-meter mark in 75.33 seconds, but due to the attack, a delay of 11.43 seconds was observed.

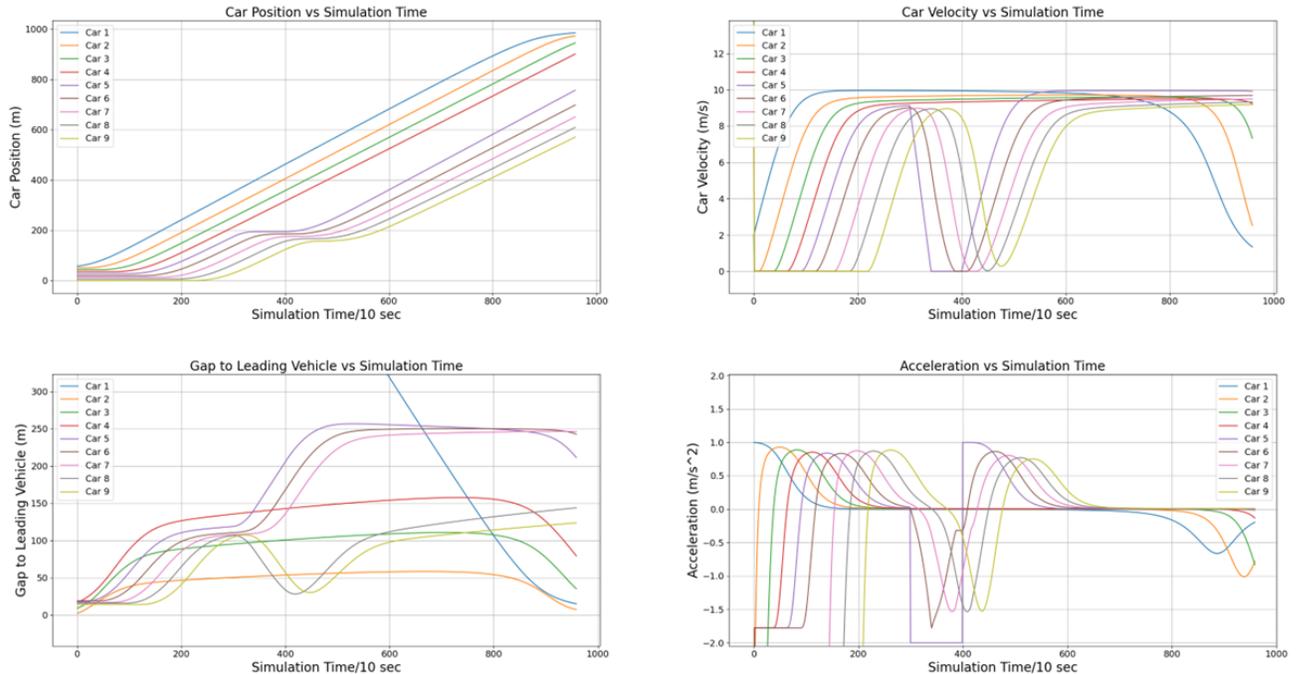

**Figure 6. CAV Platoon under Acceleration Manipulation Attack**

**Attack type: Multiple Vehicle Attack**

Previously, a single vehicle was selected and was subjected to a cyber-attack. Now, three different vehicles were selected and attacked simultaneously. This coordinated attack led to significant disruption in the flow of traffic. Just 10 seconds into the simulation, vehicles 4, 5, and 8 were manipulated with false inputs. Specifically, vehicle 4 received inaccurate data, perceiving its lead vehicle as 60 meters closer than reality. Meanwhile, vehicle 5 has its acceleration tampered with, set to a fixed value of $2m/s^2$. The consequences unfolded rapidly: vehicle 5 collided with vehicle 4, 25 seconds after the attack commenced. Furthermore, vehicle 8 encountered an attack that rendered it incapable of recognizing any vehicles ahead. As a result, it assumed the role of platoon leader and accelerated, leading to a collision with vehicle 7 within just 5 seconds of the attack's initiation. This disruption caused by this attack on traffic dynamic is portrayed in Figure 7.

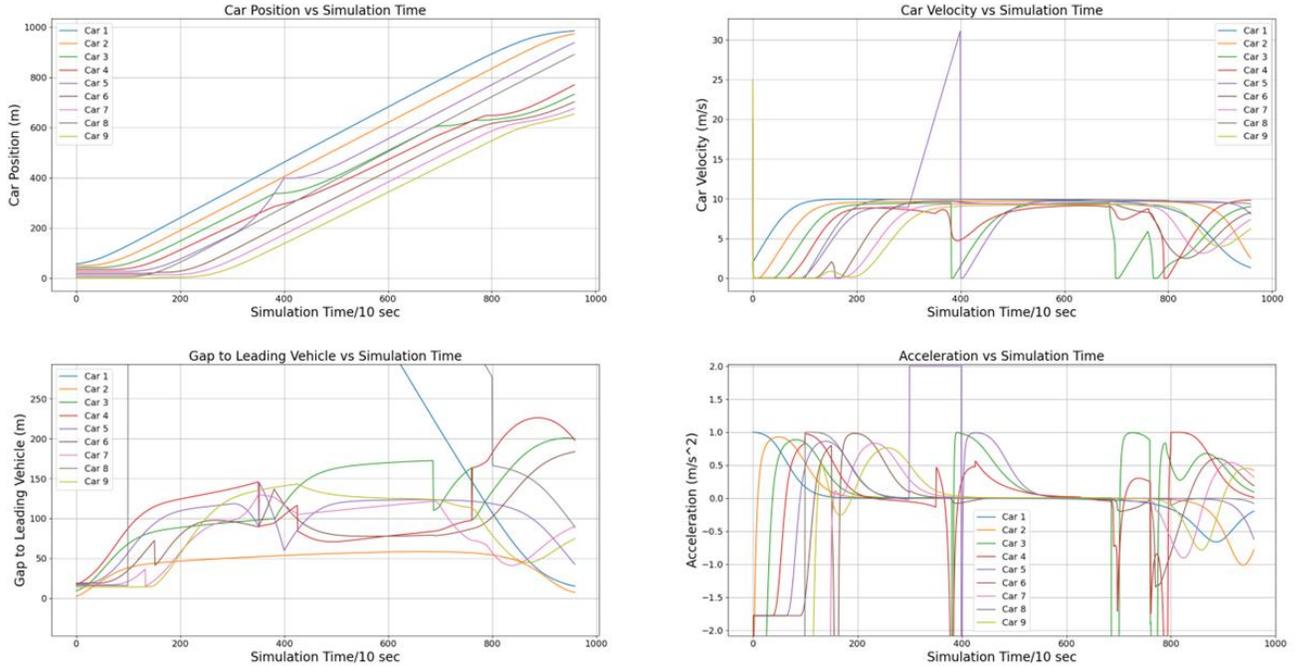
**Figure 7. CAV Platoon under Multiple Vehicle Attack**

## 4. Conclusions and Discussions

CAVs bolster many advantages ranging from reduced congestion, and safer traffic to reduced fuel consumption, making it the goal for the future mode of transportation. However, it is very vulnerable to cyber-attacks. This paper aims to investigate various potential cyber threats and their repercussions on the traffic flow. Focusing on the worst-case scenario where a vehicle's inbuilt sensor and wireless communication are compromised, the study showcases the consequences of such attacks- collision and delay.

In this study, we explored attacks in two scenarios: when a single vehicle is targeted and when multiple vehicles are subjected to attacks. The outcomes were notably more severe when multiple vehicles were affected. Among the instances where a single vehicle was targeted, those inducing acceleration were identified as particularly dangerous, leading to rear-end collisions with the vehicle ahead. On the other hand, attacks causing the targeted vehicle to decelerate didn't result in collisions but extended travel time. This delay markedly exceeded the normal duration expected without any attacks. Although CAVs are envisioned as a safe and efficient mode of travel, these cyber threats pose a great challenge that needs to be tackled and overcome.

In this paper, we have tested an innovative car-following model for connected vehicles where the subject vehicle selects acceleration with the safest expected outcome. Through our analysis, a crucial finding was identified: attacks causing a vehicle to decelerate had a milder impact compared to those inducing acceleration. When a vehicle slowed down, its followers also decelerated, aligning with the principles of the car following model wherein subsequent vehicles adopt the minimum calculated acceleration. Conversely, vehicles that accelerated ultimately collided with the vehicle ahead.

One of the primary limitations of this study is that the subject vehicle is only concerned with the leading vehicles. Establishing communication with the trailing vehicles could mitigate the accident risks. The velocity and position data of the attacked vehicle can help the leading vehicle to either accelerate or exit the platoon and give way, to avoid collision.